\documentclass[aps,twocolumn,tightenlines]{revtex4}
\usepackage{amsmath}
\usepackage{amssymb}
\usepackage{MnSymbol}

\usepackage{graphicx}
\usepackage{bm}
\newcommand{\be}{\begin{equation}}
\newcommand{\ee}{\end{equation}}
\newcommand{\ba}{\begin{eqnarray}}
\newcommand{\ea}{\end{eqnarray}}

\begin{document}

\bibliographystyle{apsrev} 

\title {Possible Role of the WZ-Top-Quark Bags 
 \\ in Baryogenesis   }
\author{
Victor V.Flambaum \\
%
School of Physics, University of New South Wales, Sydney\\
  2052, Australia\\
%
Edward Shuryak,\\
%
Department of Physics, State University of New York,\\
 Stony Brook, NY 11794, USA}

    \date{\today}


    \begin{abstract} 
 The heaviest members of the SM -- the gauge bosons $W,Z$ and the top quarks and antiquarks --
may    form collective bag-like excitations of the Higgs vacuum provided their number is 
    large enough, both at zero and finite temperatures. Since Higgs vacuum expectation value (VEV) is significantly modified
    inside them, they are called ``bags". 
    In this work we argue that  creation of such objects can explain certain numerical studies of cosmological baryogenesis. 
   Using as an example
    a hybrid model, combining inflationary preheating with cold electroweak transition, we
 identify  ``spots of unbroken phase" found in numerical studies of this scenario with such $W-Z$ bags. 
   We argue that the baryon number violation should happen predominantly inside these objects, and  show that the rates
calculated in numerical simulations can be  analytically explained 
using finite-size pure gauge sphaleron solutions,
developed previously in the QCD context by Carter, Ostrovsky and Shuryak (COS).
 Furthermore, we point out significant presence of the top quarks/antiquarks in these bag
(which were not included in those numerical studies).   Although the basic
 sphaleron exponent  remains unchanged by the top's presence, we find that tops help to   stabilize them
for a longer time. Another  enhancement of the transition rate comes from the ``recycling" of the tops in the topological transition.
Inclusion of the fermions (tops) enhances  the sphaleron rate
by up to 2 orders of magnitude. We finally discuss  the magnitude of the CP violation needed to explain the
observed baryonic asymmetry of the Universe, and give arguments that the difference in the top-antitop population in the bag of the right magnitude
can arise both
from CP asymmetries in the top decays and in top propagation into the bags, due to Farrar-Shaposhnikov effect.  
    \end{abstract}
    
   \maketitle 
    

    \section{Introduction}
\subsection{The Baryogenesis}
  The question how
 the observed baryonic asymmetry of the Universe was produced is 
  among the most difficult open questions of physics and cosmology. Sakharov \cite{Sakharov}
 had formulated three famous necessary  
conditions:  the baryon number and the CP violation, with obligatory deviations from thermal equilibrium.  
Although all of them are present in the Standard Model (SM) and standard Big Bang cosmology,
the known mechanisms creating the baryon asymmetry can only produced
effects many orders smaller than the observed amount, usually expressed as the ratio of the baryon
 density to that of the photons
  $n_B/n_\gamma\sim 10^{-10}$.  To find a scenario in which this puzzle can  at least in principle be solved
  is one of the motivation of this paper.
  
   It follows  from general anomaly relation and, in particularly, from known electroweak
   instanton and sphaleron solutions, that baryon number can be violated by certain processes
   providing a change in   topology of the gauge field. In the broken phase 
   of the SM we live in today these processes are basically frozen, by a 
   very high barrier associated with Higgs VEV. 
 On the other hand, in the symmetric phase -- the electroweak plasma -- the sphaleron rates are suppressed only by the powers of 
 the coupling:  those
  rates 
 are high enough to wipe
out any primordial asymmetry, if it has been generated cosmologically prior to
the electroweak transition era. 
These consideration force us to search for resolution of the  baryon asymmetry puzzle
at narrow temperature interval at or right below the electroweak phase transition.
For review of the field see e.g. \cite{RS,leptogen,Dine:2003ax}.

The bubbles associated with the expected 1-st order electroweak phase transition had attracted significant attention
in 1980's-1990's.
 Moving walls of such bubbles may supply 
the necessary local deviations from thermal equilibrium. The rate of 
thermal sphaleron transitions, both in equilibrium  and near the bubble walls,
 has been studied  extensively, see \cite{Klinkhamer:1984di,Arnold:1987mh,Carson:1990jm,Moore:1995jv} and many more.
However, as the experimental limits on the Higgs mass have over time evolved upward,  to $M_H> 116\, GeV $ today, it became 
clear that the first order transition is actually impossible in the SM. (Transition is a crossover for $M_H> 80\, GeV $, see again \cite{RS} for  details and the original references).  
After this has fact has been acknowledged, people either looked at various phenomena $beyond$ the
SM such as its supersymmetric extensions, which may still allow some window for the
1st order transitions. 
 
An alternative
  is the modification of the standard cosmology.
Instead of the usual Big Bang scenario, with its adiabatically slow  
crossing of the  phase transition and  tiny deviations from equilibrium, 
the so called ``hybrid" scenario  has been proposed
\cite{GGKS,KT}, see also \cite{Felder:2000hj,GarciaBellido:2002aj} and many subsequent works.  
It combines the end of the inflation era with the establishment of
the  electroweak broken phase.  In it
 the nonperturbative production of long-wavelength gauge and Higgs fields
 happens far from equilibrium, returning to
  equilibrium   later,   at a temperature
 well
 below  the critical temperatures  $T_c$ of the electroweak transition; therefore it is sometimes  called the ``cold" electroweak scenario.
 While based on some fine tuning of the  unknown physics of the inflation, it avoids many  pitfalls of the standard  cosmology, such as ``erasure" of asymmetries generated before the electroweak scale.

 In such a scenario there are coherent oscillations of the gauge/scalar fields \cite{GBG,Dima}: those 
  have been studied in detail in  real-time lattice simulations \cite{GarciaBellido:2003wd,Tranberg:2003gi}. 
  Some of these  authors focuses on the problem of generation of primordial 
magnetic fields, see e.g. \cite{DiazGil:2007dy} in the same scenario, which we will not discuss. 
The simulated models include two scalars - the inflaton and
the Higgs boson -- and the electroweak gauge fields of the
SM, in the approximation that the Weinberg angle is zero
($Z$ is degenerate with $W$). All fermions of the SM are
ignored:  the effect of the top quark in particular
is the subject of the present paper.
After inflation ends, all bosonic fields are engaged
in damped oscillations
 for relatively short time, at the end of which
the Higgs VEV and gauge fields stabilize to their equilibrium
values, with the bulk temperature  $T_{bulk}\sim 50 \, GeV$, 
well below the critical (crossover) temperature. 

In this work we will use the results of these simulations as a quantitative example,
although some our results should also be applicable for standard cosmology as well.
We start with explaining some of the
results of these simulations (the sphaleron size and rate) 
 in simple analytical models, 
 and then proceed to discuss what are changes induced by the top quarks. 

Another ingredient of the baryogenesis problem, the CP violation, had  also evolved 
during the last decades. The nature of observed CP-odd effects, first discovered in kaon decays and lately
in decays of the $B$ mesons, is rather firmly established and its origins traced to the complex phase
of the CKM matrix. It has been argued that it should enter as the so called Jarlskog determinant , with all mass differences explicitly present polynomially, leading to
effect $\sim 10^{-19}$, way too small for baryogenesis.  
The magnitude of the CP violation in the SM is however still hotly disputed: the effect must be proportional to the  Jarlskog determinant 
only if all fermionic masses appear explicitly in the numerator. This would be the case if the relevant scale of all loops is much larger 
than all the masses.
This is not the case e.g. in the famous CP violation in the kaon decays.
It has been argued recently \cite{Hernandez:2008db} that next-to-leading order dim-6 operator appears which contains $1/m_c^2$ in the coefficient: if so,
numerical study of the possible role of this operator \cite{Tranberg:2009de} found CP effect even 4 orders of magnitude larger than needed! We will return
to critical discussion of this issue at the end of the paper. 

 Vast
literature exists on possible leptogenesis scenarios,  generating the asymmetry  in the
leptonic sector, for overview see e.g.  \cite{leptogen}. If this is the case however, leptonic asymmetry should  still be
converted into the baryonic asymmetry, again 
  at or near the electroweak transition scale. We will not discuss this option in this paper, together with many other
  ideas beyond the SM.

\subsection{The $W-Z-top$ bags}
   \label{intro_bags}
Generic argument for multi-particle states 
 due to Higgs attraction is that however weak the forces can be, if  it is an universal attraction,
a  sufficiently large  number $N>N_c$ of particles it will become very strong. For example, an 
 extremely weak gravity force binds the planets and stars. 
 This generic argument however assumes that Higgs is light enough, so that sufficiently many particles can be collected
 inside the region of a size $< 1/M_H$, so that the universal Higgs-induced forces are now yet cut off by its mass.
 In the context of
 Higgs attraction
 it was discussed in particularly for light quarks in
 \cite{Shaposhnikov_bags}, in which case the $N_c$ is astronomically large and it worked only for very light Higgs bosons.

More
recent round of ideas has focused, not surprisingly, on the heaviest member of the Standard Model (SM),
  the  top quarks. The first question was whether a $single$ top quark 
would significantly  modify the Higgs vacuum in their vicinity to make a ``bag"
\cite{MacKenzie:1991xg,Macpherson:1993rf}.  This idea was soon refuted (see e.g. \cite{Farhi:1998vx})
by quantum one-loop corrections.  Similarly, the idea does not work for ``realistic" Higgs mass and binary $\bar t t$ or $tt$ systems. 

 Frogatt, Nielsen and collaborators \cite{Froggatt:2008ns}  were the first to look at multi-top system. They 
provided simple hydrogen-atom-like estimates 
for a system of $N=12$ top quarks (6 top and 6 anti-top for the 
lowest $S_{1/2}$ orbital), which suggested that its binding energy can be large
and nearly cancel the mass! Unfortunately, our more accurate mean field  calculations \cite{Kuchiev:2008fd} also refuted it. The
 binding is nonzero  for massless Higgs,  although it is much smaller than estimated in \cite{Froggatt:2008ns}. With the 
``realistic" Higgs mass  we found that the 12 top-antitop system is unbound.
 Further discussion of the binding of the 12 top system can be found in \cite{Richard:2008uq}, who  refined the conclusion in the  form of the maximal Higgs mass
 at which binding occurs:
 \be  M_H < M_c(12)\approx 49 \, GeV \ee
 
Systematic study of the top-bags in relativistic regime has been started  in the 
 paper  \cite{Kuchiev:2008fd}. One-loop quantum corrections to the bags have been investigated in a separate paper \cite{Crichigno:2009kk}, 
 following basically the line of thought of Farhi et al. We have found that at Yukawa coupling corresponding to the top quark the quantum (one-loop) effect are
 still under control, for any $N$, while it is not so for hypothetical fermions few times heavier than tops.
 Significant progress has been made in the companion paper  \cite{topbags}, in which we have studied binding of various bags, filled  with bosons ($W,Z$) and fermions ($\bar t, t$).
  We have found that bosonic bags may exist even in vacuum, for ``realistic" Higgs masses $\sim 100 \, GeV$, while purely top bags do not. 
  Those bosonic bags should however have very large number of quanta, of the order of thousands, to get bound.

One reason to study the $W-Z$-top-bags is methodical: they are only the third (and highly relativistic) family of manybody bound objects, besides atoms and nuclei.
 Unfortunately their experimental production at accelerators seems to be impossible, mostly because it is highly improbable to produce many tops
 in a small vicinity from each other. In this paper we suggest another motivation for their study,
the  cosmological one. One of  the main messages of this paper is that 
metastable (mechanically stabilized) multi-quanta bags replace
   the  ``metastable bubbles" considered 20 years ago,  being the  arena in which 
 the baryon number violating sphaleron transitions (and possibly also CP violation)   takes place.

\subsection {Sphalerons with and without the Higgs }

Semiclassical description of the tunneling through the barrier, separating
topologically distinct gauge fields, is given by
  the famous electroweak $instanton$ solutions, which however leads to  extremely low tunneling
rate
per unit time and volume, normalized by the only scale of momenta, the temperature $T$, to a dimensionless quantity
\be \Gamma_{tunneling}/T^4 \sim \exp(-4\pi/\alpha_w) \sim 10^{-170} \ee

 At
finite temperatures another option appears,  the so called sphaleron
transition, which is due to a thermal excitation $onto$ the barrier.
The original electroweak sphaleron solution found by
 Klinkhamer and Manton (KM) \cite{Klinkhamer:1984di}
 has the energy $ E_{KM}$ (the height of the barrier) of about 14 $TeV$. 
 So , at  the temperatures  at the electroweak transition scale 
 $T\sim T_c\sim 0.1 \, TeV$, the corresponding Boltzmann factor
is 
 prohibitive
\be \Gamma_{KM}/T^4 \sim exp(-E_{KM}/T) <\sim 10^{-60} \ee
However, this estimate is too naive, as the parameters of the electroweak theory are strongly renormalized near $T_c$.
More accurate calculation of the  equilibrium
 sphalerons rates
 at the electroweak cross over region give much larger rates.   For recent  update see e.g.\cite{Burnier:2005hp} who estimated those in the range 
 \be \Gamma/T^4\approx  10^{-20}\ee
 including rather large preexponent calculated in
Refs. \cite{Arnold:1987mh,Carson:1990jm,Moore:1995jv}. 
Such rates are however still too  small for the solution of the  
baryogenesis puzzle.

The main  reason of the increase of the rate, from the 
KM sphaleron to revised one is of course the reduction of the Higgs  
 VEV from its vacuum value  $v=246\, GeV$ to near-zero. 
(Recall that there is  no Higgs-related barrier in the symmetric phase, and thus no exponential suppression:
but this is not a blessing since a transition rates faster than cosmological expansion leads to ``wipe out".)
 
  In the hybrid scenario one asks: what
 is the barrier height and the sphaleron transition rates for a no-Higgs-VEV bag of  a specific size $\rho$? 
Although with a completely different (QCD) motivation, the  analytic
answer to this  question 
 was already known, given by the so called COS sphaleron
\cite{Ostrovsky:2002cg}.  
 The main obstacle for finding it earlier was that, lacking 
a nonzero Higgs VEV,  classical  gauge theory 
is conformal and has no dimensional parameter. Thus the minimal
sphaleron mass can only be defined if some additional condition
is imposed,  e.g. that the r.m.s. $size$ of this object is fixed 
\be \rho^2 ={\int r^2 B(r)^2\ d^3r \over \int B(r)^2\ d^3r }\ee 
 Adding another constraint, that the Chern-Simons
number of the configuration $ N_{CS}$ is also fixed to some value, one can minimize the energy over all possible magnetic field configurations
and find a ``sphaleron path", the  set of  configurations which describe
the barrier separating one topological valley from the next. 

COS result for the energy as a function of the Chern-Simons number
 can be written in a parametric form, with the parameter $k=-1..1$
\be E_{COS}= {3\pi^2  \over g^2 \rho} (1-k^2)^2
\label{eqn_COS_energy}\
\ee 
\be N_{CS}={1\over 4} sign(k)(2+|k|)(1-|k|)^2 \ee
The $k=0, N_{CS}=1/2$ configuration is the top of the barrier --the COS sphaleron.

We will be arguing below that the topological objects found numerically
in real-time are in fact close to COS sphalerons with the total energy 
\be E_{COS}\sim 2 \,\,
TeV \ll E_{KM}\approx  14 \, TeV \ee
 much smaller
 than in the broken phase.  (Once again: it is only possible because
 they are located in the ``no-Higgs spots" with depleted Higgs VEV.) Obviously the corresponding
 Boltzmann factor is  dramatically reduced, falling into the range in which one can discuss baryogenesis.

\subsection{The aims and the structure of this paper}
  In a sentence, we will show that the presence of the
 $W-Z$-top bags should significantly enhance the rate of baryon number violating processes.

  A bit more detailed summary is that there are at least three important effects we will consider:\\
 (i) As we just discussed,
   the baryon number violation
  rates inside such bags are enhanced by many orders of magnitude due to
  the absence of the Higgs VEV and related high barrier. In other words,
  instead of KM sphalerons one should use the COS ones, reducing the barrier height from 14 to only about 2 TeV. 
  We will be able to analytically estimate the sphaleron rates, which compare
 well with numerical simulations.\\

(ii)  Accounting for top   ``bags" in this setting we find that
 tops help mechanically stabilize the bags. In a way, the metastable bags 
 play  a role similar
  to that of the bubbles (present if the electroweak transition be of the 1st order),
  namely to enhance deviations from equilibrium for rather long time.\\
  
  (iii) Last but not least,  the ``recycling" of top quarks, present in the bag, effectively
  lower the barrier further,  by about 600 GeV.   CP odd effects may lead to top-antitop population difference
  in the bags, which will result in asymmetric diffusion in the baryon number.
  
 Now about the structure of the paper. We start in section \ref{sec_bosonic} with the  discussion of the
 the results of numerical simulations  \cite{GarciaBellido:2003wd} and then qualitatively explain them
 using the COS sphaleron.
 After that, we turn to top and W bags in section \ref{sec_topbags}.
 We discuss their production and lifetime. Finally, in section \ref{sec_sphal_topbags} we will study how the sphaleron transition
itself is changed, with the account for fermions.

\section{Hybrid inflation scenario: bosonic simulations and their discussion }
\label{sec_bosonic}

\begin{figure}[t!]
\includegraphics{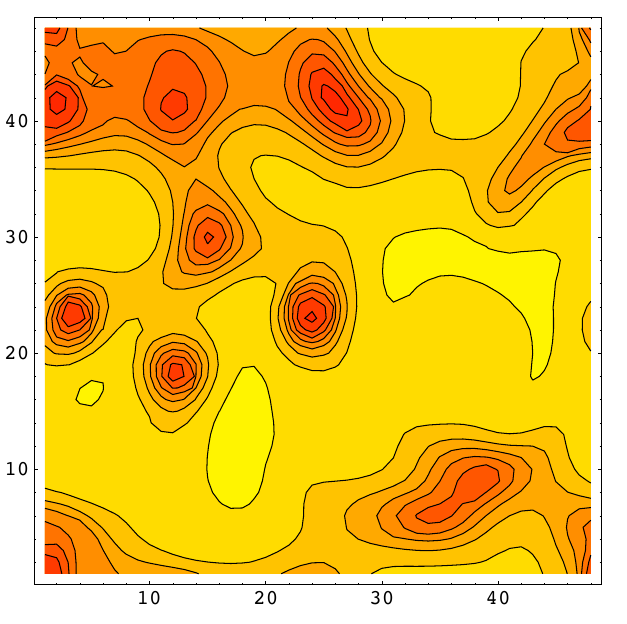}
\includegraphics{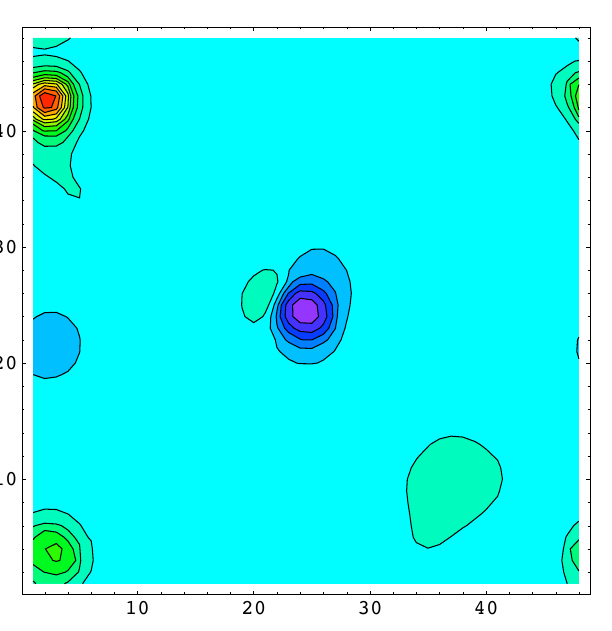}
\caption{(Color online). The
contour plots from \protect\cite{GarciaBellido:2003wd}
of the modulus of the Higgs field $|\phi(x)|^2$ (upper plot) and the topological charge density  $Q(\vec x,t)$   at 
time  $mt= 19$, for the 
model A1, $N_s=48$. Red (dark) areas in the upper plot correspond to small VEV, while yellow (light) bulk
corresponds to the broken phase. On the lower plot
 lumps of the topological charge
 density appear as red regions (dark in black and white display). While
 most of the no-Higgs spots do not have the topological transitions, all transitions seem to be inside the spots.
}
\label{contour1}
\end{figure}

(i) One important finding of the simulations is that the initial coherent oscillations of scalars soon give way
to the usual broken phase. The most important  feature is persistence of 
 ``no-Higgs spots'' 
in which Higgs VEV is very far from the equilibrium value $v$ and
is instead close to zero. The gauge fields in them have however rather high magnitude.
 Fig.\ref{contour1} (from \cite{GarciaBellido:2003wd})
show an example of a snapshot 
  of the Higgs field modulus. Typically
the volume fraction occupied by such ``no-Higgs spots"  is of the order of several percents and is decreasing with time. 

(ii) The second
important findings is that of
 topologically nontrivial fluctuations
of the gauge fields. As shown in the lower Fig.\ref{contour1}, those are well localized.
According to the anomaly relation, changes in topology leads to 
violation of $B+L$ (baryon and lepton) numbers, via emission/absorption of fermions, 12 in the SM.

It was observed 
 that topological
fluctuations happen only $inside$ 
 the ``no-Higgs spots" mentioned above. Indeed, this becomes apparent from  the distribution
of the topological charge shown in the lower part of Fig.\ref{contour1},
 for the 
same time  configuration of the Higgs as shown in the upper part of the  Fig.\ref{contour1} .
Of course, it is only one snapshot, but the authors found from the simulations that it is true for the whole sample.

The fraction of no-Higgs spots which induced   topological transitions
is also in the range of a  percent. More precise measure is the so called
 sphaleron rate is defined by the mean square deviation from zero of the Chern-Simons number 
\be \Gamma(t) = {1 \over m^4 V}{d \Delta N_{CS}^2\over dt}\ee
(here and below all quantities are defined via one characteristic mass parameter $m$: for the
simulations its value is about $264\,\, GeV$, close to $v$. For the definition see Appendix A.)

Since the process only exist for finite period of time, ref.\cite{GarciaBellido:2003wd} reports
 its time integrated value defined by some time integral
\be I(m t)=\int^t_{t_i} d(mt) \Gamma(t) \ee
during which it basically happen. (Too early there are no gauge fields and too late there are no no-Higgs spots.)
 By the end of the process ($m t < 45$) it is in the range
\be I\sim 10^{-4}\ee
for more details about different parameters sets see Table II of \cite{GarciaBellido:2003wd}. 
This quantity, as well as of course snapshots like those shown in Fig.\ref{contour1},
directly give the spatial distance between the
 topological fluctuations 
 $R_{sph}m=  20..30$.

\begin{figure}[htb]
\includegraphics[width=8cm]{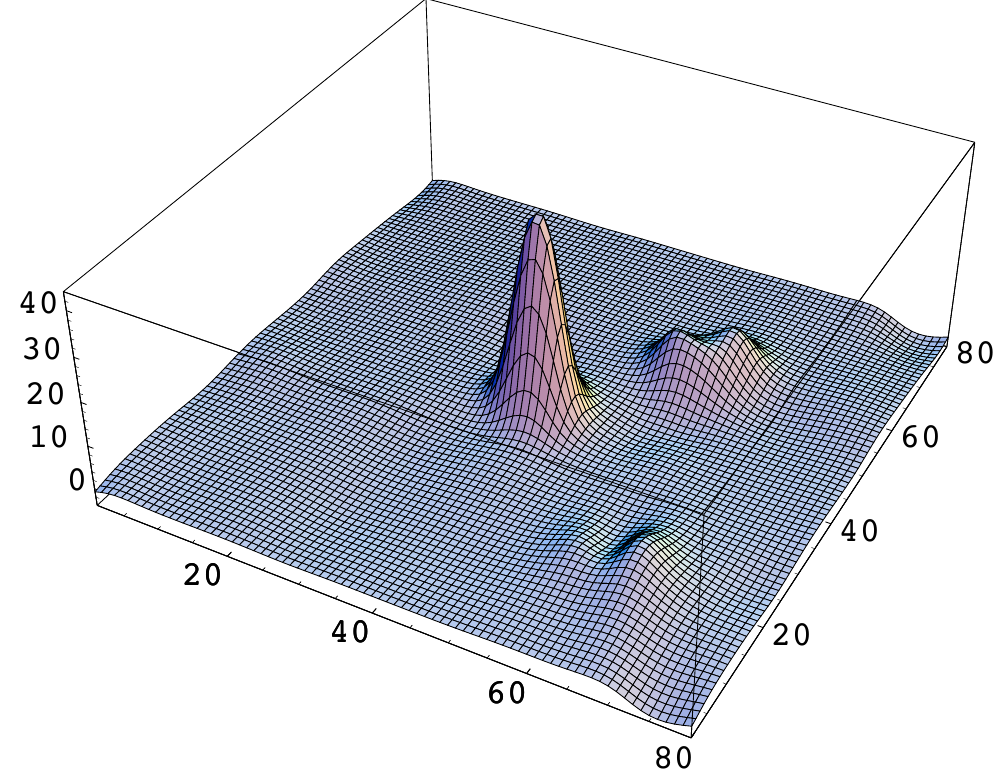} \includegraphics[width=8cm]{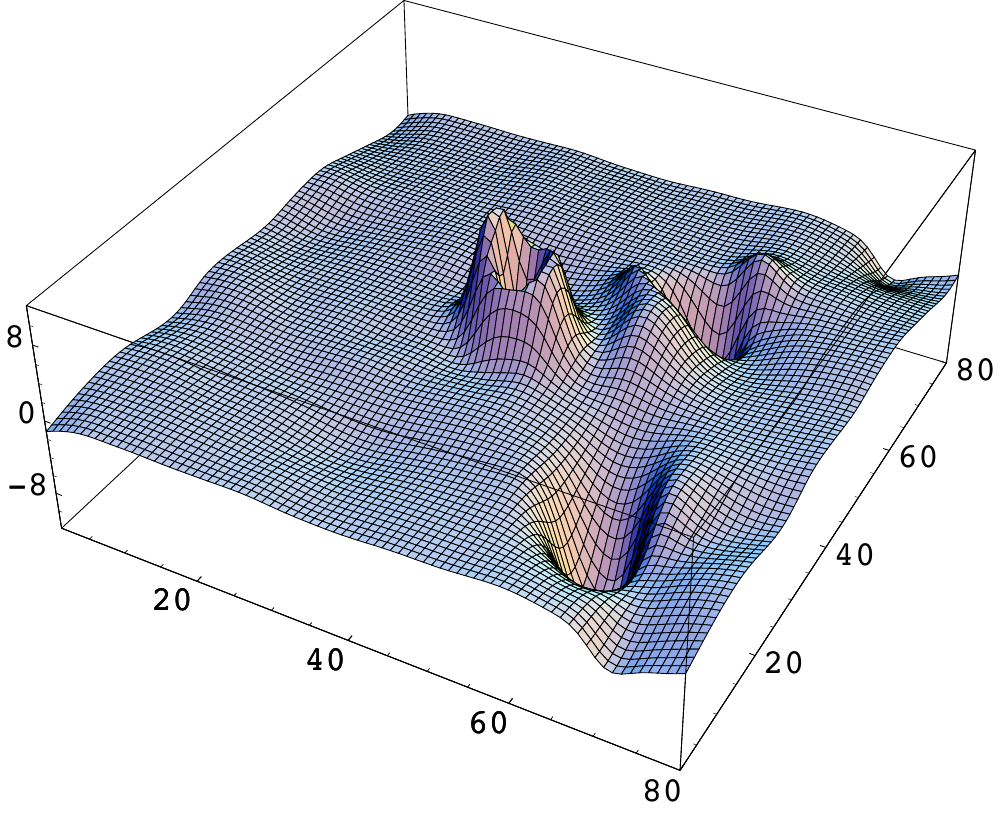} 
\caption{(Color online). From \protect\cite{GarciaBellido:2003wd}.
Two snapshots of the topological charge, at times $mt=18$ and $19$.}
\label{fig_top_explo}
\end{figure}
\begin{figure}[htb]
\includegraphics[width=8cm]{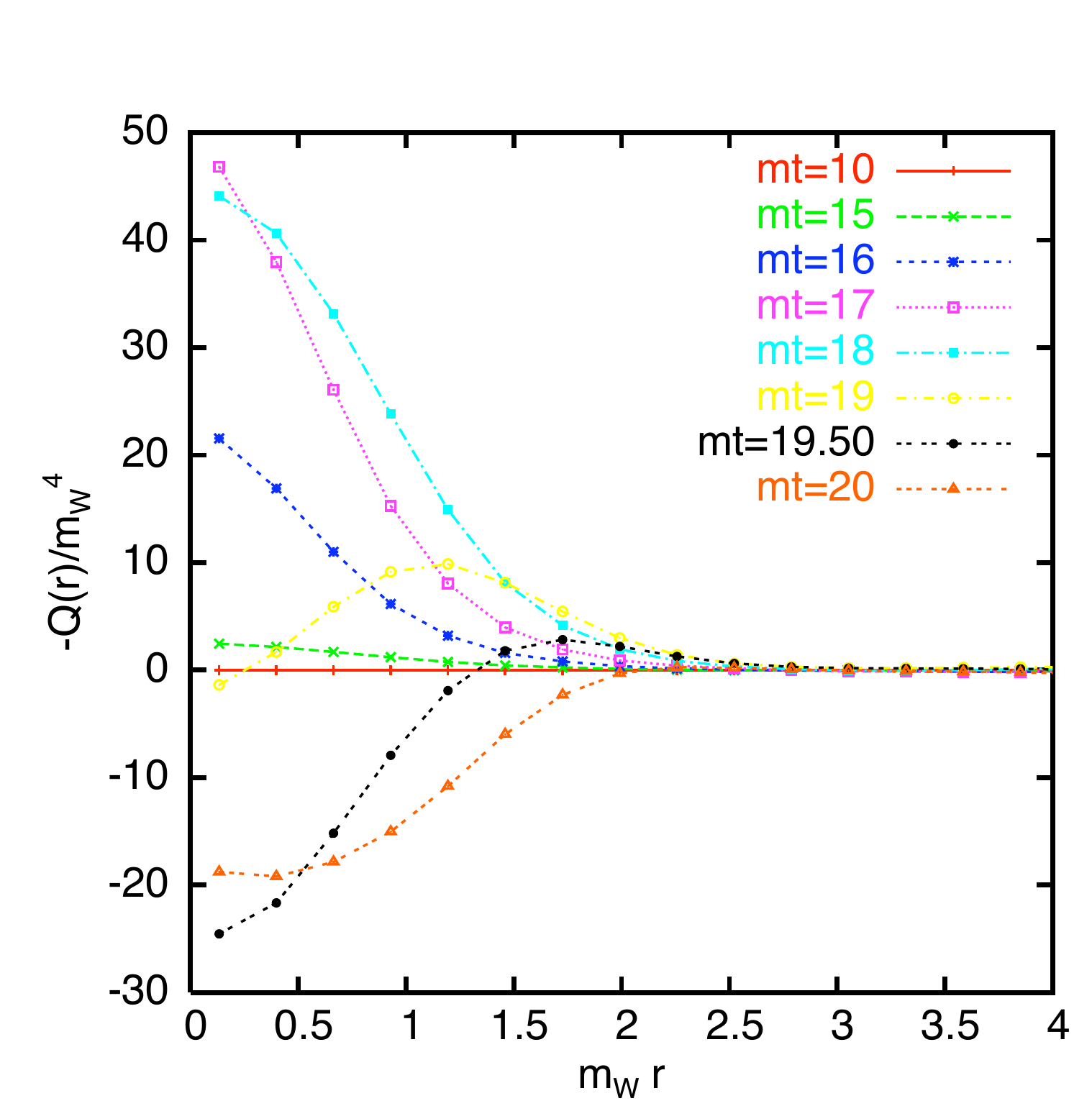}
\includegraphics[width=8cm]{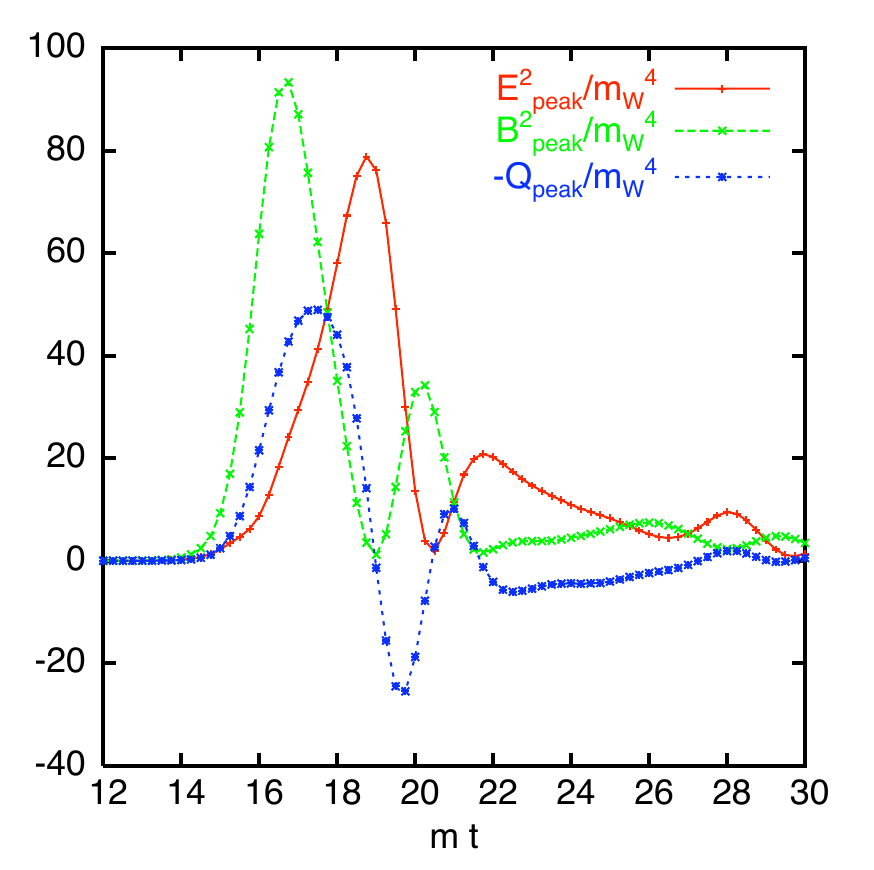}
\caption{(Color online). From \protect\cite{GarciaBellido:2003wd}. Upper: The spatial profile of the charge density peak for
various times. Lower: The value of the electric, magnetic and charge density
at the center of the peak as a function of time.}
\label{fig_profile}
\end{figure}

(iii) Space-time evolution of the  topological charge Q is shown 
in Fig.\ref{fig_top_explo} as two snapshots, and in more detail in Fig.\ref{fig_profile}
for the maximal $B^2(t),E^2(t)$ and $Q(t)$, for one particular 
fluctuation. One can see that the fluctuation at some time moment 
is very much concentrated in a small spherical cluster (the upper one in Fig.\ref{fig_top_explo})
is followed by an expanding spherical shell (the lower one in Fig.\ref{fig_top_explo})
which gets near-empty inside. 

The decomposition into electric and magnetic components of the field
is shown in Fig.\ref{fig_profile}. One can see that the fluctuations starts as nearly (90\%) 
magnetic object at $mt\sim 17$,  with the electric field
and thus the topological charge $\sim \vec{E}\vec{B}$
peaking some time later. Then there appears an expanding
shell, followed by the magnetic field  rebounding  to its secondary
maximum of smaller amplitude. We will return to discussion of all those features in the next chapter.

Some interesting findings of those studies we will not discuss but yet would like to mention. In Ref.\cite{half_knots} it has been shown
that ``hot spots" have some topology in terms of the Higgs winding number, and that their favorite values is about 1/2 (thus named half-knots).
There is some correlation between this winding number and $N_{CS}$ (and thus the sphaleron formation) which deserves to be  studied further. 

Before modelling these numerical results, let us summarize  
what is it exactly we would like to achieve:\\
(i) to model the structure of the ``no-Higgs spots" by WZ-top bags\\
(ii) to model the topological fluctuations in them by the COS sphalerons,
 explaining their structure,   size and ultimately the rate\\
 (iii) to discuss specifically the modifications coming from the top quarks, as those were not included in numerical simulations.

\subsection{``No-Higgs spots" as  the $WZ$ bags}
In Ref.\cite{topbags} we have made some first steps
toward understanding of the states  of the gauge bosons in certain 
 spherically symmetric Higgs backgrounds. 
Our main finding was that while
outside of the bag the gauge quanta $W,Z$ are massive and have 3 polarizations, in the bags
the longitudinal component obtains additional strong repulsive potential. It also is mixed with electric polarization, which is indirectly affected
by this potential.
Only magnetic one remains unaffected by this repulsion: their properties were documented in \cite{topbags}.

In principle, at finite $T$ one needs a complete set of $W,Z$ states for the partition function of the bag. Furthermore,
 in order to model the  ``spots" appearing near reheating, one should also do it out of equilibrium, which is even more complicated.
%
However, since such states are not yet known in full, we will start here with two
grossly simplified limits, assuming
that those spots contain so large
number of W that they can approximately be treated macroscopically.

 Thermodynamically
stable bubbles of the symmetric phase can only coexist with  the bulk in the broken 
phase only $at$ the  transition temperature $T_c$, provided the phase
transition is 1st order.  As we already mentioned in the Introduction,
such option has been excluded by the combination of experimental limits on Higgs mass and
lattice studies of the transition.  And we also consider a ``cold model" in which the equilibrated bulk
is at a temperature well below the critical one, $T<T_c$. Therefore,  we discuss the objects which are not thermodynamically but only $mechanically$
stable.

What we consider here are  a very simple variant of the bag model. Let us assume that
rescattering are rapid enough to make kinetic equilibrium inside the bag. There is no
$chemical$ equilibrium, as the number of $W$s produced are determined by early time dynamics:
so we need to introduce both the internal temperature $T_{in}$ and the fugacity $\xi_W=exp(\mu_W/T_{in}$ . Rescatterings
will force the internal momentum distribution to be thermal
\be N_W=V_{spot} g_W \int {d^3k \over (2\pi)^3}{1 \over \xi_W^{-1} exp(\epsilon(k)/T_{in})-1}   \label{eqn_thermal} \ee
 even if  there is no chemical equilibrium, say the number of gauge degrees of freedom can be different from the 
 maximal number $g_W=6$ (e.g. only  magnetic modes excited, without electric or longitudinal
 ones, as we discussed in the companion paper) or $ \xi_W$ different from 1.

Condition of mechanical stability we will write as
\be p_{in}=B+p_{bulk}  \ee with the bag constant $B$ created by the Higgs potential, assuming zero
VEV inside
\be B=\lambda v^4/4 =m^2 v^2/4\ee
(loop and T-dependent corrections can be easily included but are ignored.
All other degrees of freedom of the SM such as
 quarks and leptons are ignored here because they have not yet been produced.)
 Using further Boltzmann approximation (ignoring 1 in denominator of  (\ref{eqn_thermal})) and also ignoring
 $p_{bulk}$ for the estimate, one gets the mechanical stability condition in a form
\be B = g_W \xi_W{\pi^2 T_{in}^4 \over 45} \ee
 or  the internal temperature
 \be   T_{in}\xi_W^{1/4}({g_W\over 6})^{-1/4}= .66 m\approx 174\, GeV\ee
So, if it is an equilibrium W gas (all factors are 1) the internal temperature
inside the spot is indeed well  above the electroweak critical temperature $T_c\sim 100\, GeV$,
making it a spot of symmetric phase.

Since the $W$ bag is only mechanically stabilized, it is relaxing by cooling and thus disappear relatively quickly.
Indeed, this is what happens
in simulations, giving the characteristic lifetime of the no-Higgs spots
\be \tau_{hot\,spots}\sim 20/m\ee 
It is however important for the sphaleron rate that -- in the bag approximation we use --
 the  inside temperature $T_{in}$ is $not$ changing while this bag shrinkage takes place, 
because it is related with the Higgs bag constant.

\subsection{ The  sphalerons, their structure and  size}

Trying to understand the results of these  numerical simulations, one can
ask in particular the following questions:\\
 (i) How so high-magnitude
gauge field  can be  produced? Why does it happen only inside the no-Higgs spots?\\
 (ii) What is the energy needed for topological fluctuation? What is the gauge field structure? 
 Can one understand its subsequent evolution and decay products?\\
(iii) Can one estimate the rate of their production?

 The fact that topological transitions  happen only inside the
no-Higgs spots is rather simple to understand: as we already mentioned in the introduction
in the broken phase the height of the barrier (the mass of KM sphaleron)
is prohibitively large. In a spot the height is not zero, however, as it has finite size $\rho$.
As we already mentioned, it is given by the mass of the COS sphaleron $\sim 1/\rho$ (see (\ref{eqn_energy})).

 In a scale-invariant 
classical YM theory the size  $\rho$ is an arbitrary
parameter.
In a Big Bang however ( as well as in numerical models we discuss) this size
is to be determined by the optimal scenario maximizing the transition rate.

Although the bag production is a result of a complicated nonequilibrium process, it is reasonable to 
assume that large-size bags are cut off exponentially in bag volume.
Furthermore, we model it by a thermal fluctuation with the $inside$ temperature. In a simple bag model
the pressure is $p=p_{in}(T)-B$ while the energy density is $\epsilon=\epsilon_{in}(T)+B$. Mechanical stability condition
for the total pressure of the bag $p=0$ (more accurately, the outside pressure in the r.h.s., which is very close to zero ),
or $p_{in}(T)=B$. Furthermore, if the inside of the bag is filled by massless gas of any kind, $ \epsilon_{in}(T)=3p_{in}(T)=3B$
and thus the energy of the bag is $4BV$. Since our bags are not asymptotically large, perhaps some coefficient (to be later determined)
can be used instead: so we finally write
 the bag Boltzman factor  as $exp(-\beta BV)$

So, 
 the optimal size $\rho_*$ is the extremum 
 of the following expression
\be \Gamma_{sph}(\rho)\sim exp\left[-
{ \beta B} ({4\pi \rho^3 \over 3}) - {3\pi^2 \over g^2 \rho T_{in}} \right] 
\label{eqn_cos_rate}
\ee 
 Here the first term schematically represents the probability to create large bags, while the
 second is the Boltzmann factor for the COS sphaleron, with the mass from (\ref{eqn_COS_energy}).  
  The  resulting optimal spot size is then 
 \be    \rho_*= \left(  { 3\pi \over  4\beta  T_{in} g^2 B }    \right)^{1/4}  \label{eqn_optimal_size}
 \ee  
 $independent $ on the degrees of freedom filling the bag.
 Comparing it with the ``numerically observed" $\rho m \sim 3.9$, Figs . 2 and 3, at the moment of maximal $B$, 
 one may extract the value of the parameter $\beta$ which turns out a bit smaller than $1/T_{in}$.
We repeat that the model is crude and we only need it to discuss later the influence of the tops on the whole picture.

It is useful to discuss at this point how semiclassical are those sphalerons.  Note that the
  parametrically large quantity here is electroweak coupling $4\pi/g^2=1/\alpha_{ew}$,
  which determines that the sphaleron consists of many gauge quanta and thus can be
  treated semiclassically. As the radius contains its power $1/4$ we get only factor 2 from it.
  The number of gauge bosons involved can be estimated from the action/$\hbar$ which  is about $O(10)$.
  It is perhaps still large enough for the semiclassical analysis we use, but -- as we will see later -- not large enough
  to ignore the back reaction from the 12 fermions associated with the electroweak anomaly.

 \subsection{ The shape and field structure of the sphaleron}

 Ideally the sphaleron transition may proceed with
 the total energy of the gauge quanta exactly equal to to the height of the barrier.
 In this case at 
 the time the system reaches the maximum -- the ``sphaleron moment''
-- the kinetic (electric) energy is zero, after which the system may fall downhill
into a topologically different configuration. 

An interesting option pointed 
 out recently by Kuchiev \cite{Kuchiev:2009zm,Kuchiev:2009rz}  is that the optimum energy may be $less$ than the height
 with tunneling through the barrier's top. 
 That however cannot happen in classical simulations
 we discuss.

 One may think there would be   some extra 
energy needed to fall over the top in those simulations. And indeed, as 
 seen in
 Fig.\ref{fig_profile}, this is the case. 
 The time evolution
of the magnetic and electric fields have their
``sphaleron moment'', defined as 
 the maximum of $B^2(t)$. One indeed finds that
 the kinetic-to-potential $E^2/B^2$ energy ratio  
 is indeed small at this moment, of the order of percents.

Does the magnetic field fits well to COS sphaleron solution?
The profile of the the magnetic field in  COS configuration is given by the following simple expression
 \be B^2(r)= {48 \rho^4  \over g^2(r^2+\rho^2)^4} 
\label{eqn_profile}\ee
which (unlike the KM one) is just spherically symmetric. This form does indeed
fit well the observed shape of the $B^2$ at the sphaleron moment.
The maximum of $B^2/g^2$ can be related with its radius,
yielding $m \rho\approx 3.9$, which is very close to optimal size we got above.
This value corresponds to
the total energy of the COS sphaleron
\be E_{tot}=3\pi^2/g^2 \rho \approx 2\, \,TeV
\label{eqn_energy}\ee
As we already mentioned in the Introduction, it is 7 times less than the KM sphaleron mass, and for the temperatures we are dealing with
$T_{in}=200-100\, GeV$ it makes a huge difference for the rate.

The sphaleron Boltzmann factor can now be estimated, as we know both
the total energy and the internal temperature $T_{in}$ from our bag model 
for the no-Higgs spot:
\be exp(-E_{tot}/T_{in})\approx exp(-2000\, GeV/174 \, GeV)\approx 10^{-5}\ee
This is not yet the end of the estimate,
since semiclassical sphaleron rate has also a significant
preexponent. It has not been calculated for COS sphaleron yet, so we use the KM one
\be  {\Gamma \over V m^4}\approx{\omega_- T^3 \over 2\pi m^4} N_{tr} N_{rot} ({\alpha_w \over 4\pi \alpha_3^3})^3   exp(-{E_{tot}\over T}) \ee
which includes the unstable frequency $\omega_-\approx 2M_W$, as well as the numbers due to translational modes
$N_{tr}=26$ and rotational modes $N_{rot}\approx 5300$.  There are also   factor $\sim O(1)$
from the non-zero mode determinants. We used here $\alpha_3=g_3^2/4\pi=\sqrt{2M_W/g_w^2 T}$.

Combining all the factors we find that numerical value of preexponent and exponent
nearly cancels out, leaving crudely \be \Gamma / V m^4\sim 10^{-1} \ee with accuracy
say an order of magnitude or so. With that accuracy it agrees with
 the results of the simulations which
 also finds that the number of sphaleron transitions per spot  is indeed  about several percents.

   What should happen after the sphaleron moment?
 Sphaleron decay  is classical rolling of the classical (high amplitude) gauge field downhill, from the (sphaleron) top into
 the next classical vacuum . This process was extensively studied numerically for KM
 sphaleron. Remarkably, an $analytic$ solution of the time-dependent  explosion
of  COS sphaleron has also been found, see  the COS paper.
We will not describe the details of that here, just give
the expression for the late time $t\gg \rho$ profile of the energy density of the expanding shell. 
\be 4\pi \epsilon(r,t)={8\pi \over g^2 \rho^2 r^2} (1-k^2)^2\left({1\over
  1+(r-t)^2/\rho^2 }\right)^3  \label{eqn_shell}
\ee

 Comparing the explosion of COS sphaleron with numerical data
  one can see both the similarities and the differences between them.
   As seen from Fig.\ref{fig_profile}, there is an empty shell formation at some time. However
  the inside of the shell does not remain empty: in fact the topology and magnetic field
 have a secondary peak (of smaller magnitude). Qualitatively it is easy to see why it happens.  
The COS sphaleron is a solution exploding
 in zero Higgs background, with massless gauge fields at infinity.
 In the numerical simulations we discuss  such explosion happens inside
  the  finite-size cavity. As the gauge bosons of the expanding shell hit the
  walls of the no-Higgs spot, they are massive outside.  With some probability they
 get reflected by this wall back. This is the reason why a
 secondary splash (the second magnetic peak) is
 produced. It would be possible and interesting to study this problem
 separately, obtaining COS-like explosion in a spherically symmetric bag: we hope to do so elsewhere.
  
  Let us make a small theoretical digression here. For both for KM and COS sphaleron explosions there was well known controversy 
 about relation between the Chern-Simons number and the baryon charge for non-static solutions.
 At time going to infinity the  Chern-Simons number 
 during the sphaleron explosion  has been found to be different from naively expected  $ \Delta N_{CS}=1/2$.
 This happens because even a weak field at late time still preserves some
 topology. Nevertheless, it was somewhat troublesome before 
  the issue of the fermion number violation
   has been resolved by Shuryak and Zahed in \cite{Shuryak:2002qz},
who found the analytic solution of the Dirac eqn in a  time-dependent exploding COS background.
They found that any fermions starting from the sphaleron zero mode are indeed excited
into  
physical (positive energy) states at late time. 
The outgoing fermions have simple momentum spectrum
\be {\bf n}_L (k)  = \rho\,(2k\rho)^2\,e^{-2k\,\rho}\,\,.
 \label{eqn_q_spectrum}\ee
 which is close to thermal with $T_{eff}=2/\rho$, and the average quark energy $<E_q>=3/\rho$. 
Thus the amount of baryon number
violation by sphaleron transition is in fact as expected from the anomaly.

\section{The role  of the top quarks}
\label{sec_topbags}
\subsection{Top quark production and concentration in the bags}

Since the original numerical simulations have included the gauge fields but ignored fermions, 
we have to discuss first, at quite qualitative level, what their effects can be,
relative to that of the gauge fields.  Those tops  would be added to 
 the  metastable bubbles of the symmetric phase, the no-Higgs spots,  like  the W discussed 
 above.  However, there are at least three important differences:\\
 (i) their fugacity is expected to be larger, due to more rapid production rate of $t$ relative to $W$\\
 (ii) the degeneracy factor  $ g_{\bar{t},t}=12 $ is twice larger than for gauge bosons ($6$).\\
 (iii) their binding to Higgs bags is several times larger than for $W$, which  makes it easier to  satisfy
the  mechanical stabilization of the bag. (For details on that see \cite{topbags}.)

The first question is
at what time and how the top quarks would be produced. 
Top quark has the largest coupling to Higgs: so it is produced first via
 $HH \rightarrow \bar{t}t$ process. 
 Let us crudely compare its rate relative to that of  $HH \rightarrow WW$ 
 \be {\Gamma( HH \rightarrow \bar{t}t) \over  \Gamma( HH \rightarrow WW)}\sim
 ({m_t\over m_W})^4\sim 20
  \ee
where in the first estimate follows from the fact that the (vacuum) masses are proportional to 
the corresponding coupling constants, which appears in the power 4 in the rate.
This simple estimate suggests that top production may
be by about an order of magnitude more rapid than that of $W$s, and thus they would in fact
dominate  in the formation of ``spots". 

 Obviously, specific simulations are needed to test top production rates.
We are not currently able to evaluate top  production, give out-of-equilibrium and from strongly fluctuating
Higgs fields at the preheating stage. As a crude guess one may look at an
equilibrium  density in the bulk $before$ the appearance of broken phase VEV
\be n_{max}(\bar{t}t)= {3g_{\bar{t},t}\over 4\pi^2} T^3 \approx 1.1 T^3\ee

  The next step  is concentration
of top quarks into the bags. The simple reason this is happening is that they are massive 
in the broken phase and massless in the symmetric one.  As discussed in detail in \cite{topbags}, their binding can be as large as 100 $GeV$ or so.
 Less trivial
reason is that a  ``1d kink" of the Higgs field (a 2-d surface where Higgs VEV crosses zero) possesses
fermion zero modes. In practice it means that top quarks 
can glue themselves to the $\phi=0$ surfaces and  be transported along them, eventually into the remaining
island of the symmetric phase. The binding energy in such case is nearly all the top mass, 172 $GeV$, several time the bulk
temperature. It indicates that the effect can be rather robust: but only dedicated simulations can decide how effective this mechanism 
can actually be, and what fraction of the produced tops end up in the bags.

In order to see possible degree of such concentration,  note that after tops are collected into the bags,
they remain effectively massless  their, so their density inside is given by the same expression, with high $T_{in}\sim 200 \, GeV$
needed to balance the bag pressure. 
The  density in the bulk is given by the integral with the top mass and low $T_{bulk}\sim 50 \, GeV$
\be n_{bulk}= g_{\bar{t}t} \int_M^\infty {dE E \sqrt{E^2-M^2}  \over 2\pi^2} exp(-E/T)\ee
which is smaller than the density $n_{max}$ inside the spots by about factor 300!  It is thus clear, that after the broken
phase is established in the bulk, most ($\sim$ 99\%) tops should either be annihilated or collected into the bags.

Let us first estimate it from above, assuming that $all$ heavy quarks can be collected
into the remaining spots of the symmetric phase. If so, the number of tops per spot is just
the ratio of their densities
\be N(\bar{t}t) ={n(\bar{t}t) \over n_{spots}}\sim 10^2 ({T \over m})^3 \ee
where $T$ is a characteristic temperature of the Higgs field at early time, $T/m\sim 1$.
Thus we expect $N_{t+\bar t}$ of the order of a
hundreds of tops+antitops be collected into
a no-Higgs spot.

Since we are not able to model the process quantitatively, in a highly fluctuating Higgs background we will introduce a parameter, normalizing this process to that of the W.
So, we will denote the number of $tops$ normalized to $Ws$ which made into the ``spots":
\be \kappa= {N_t^{spot} \over N_W^{spot}}\ee
The population of the spots including W, tops and antitops would then be
$(1+2\kappa)$ times larger than in bosonic simulations.

\subsection{ The $W$ and the top quark  lifetime  in the bag}

The lifetime of the $W$ according to their weak decays is, in our units, \be {\tau_W m}\sim {m \over \Gamma_W } \approx 50 \ee
The lifetime for bound $W$'s is presumably reduced, according to their total energy, increasing by another factor 2 or so.
This timescale is several times larger than the lifetime of the  ``bags" in the numerical simulations, which is of the order of 10-20
in the same units.  

Thus the weak decays can be neglected (as they were in the numerical simulations) 
and the observed lifetime in the simulations should be attributed to evaporation
of the $W$'s into the bulk, by thermal excitation. Indeed, the ``hot spots" of the reduced Higgs VEV
were considered there as mere fluctuations of the fields.  We however look at them as metastable
objects, which can come into mechanical  equilibrium and decay more slowly due to heat transport as well as
strong and weak interactions, ignored so far.

As discussed in \cite{topbags} in detail, we found that tops tend to be on the surface of the bag, due to the
influence of the ``zero mode" phenomenon. 
For how long would $\bar t,t$ quarks remain in the bag, before their decay?  Because of hierarchy of the
 couplings, one should first
look at
annihilation via
strong interactions. One of such processes leads to
 production of $other$ quarks
 via  $\bar t t- \rightarrow \bar q q$, another
 is the annihilation into gluons
 $\bar t t- \rightarrow  gg $, which give 
 the lifetime $\tau_{\bar{t}t}\sim 200/T_{in}$,
 see Appendix B.  This is about an order of magnitude longer time than
 the lifetime of the bosonic ``no-Higgs spots" in \cite{GarciaBellido:2003wd}.
 
 (Another process $\bar t t- \rightarrow HH$ has a bit smaller coupling and less final states,
 especially in color, so it is clearly subleading.)

After strong annihilation of pairs, the $remaining$ number
of tops $or$ antitops are due to random charge fluctuations during the formation stage 
 \be N_{t/\bar t}\sim \sqrt{N_{t+\bar t}}\ee

Assuming that this is the case, one may ask how the tops would decay further.
 In vacuum the physical top quark
is much heavier than gauge quanta,  so they decay  $t\rightarrow b+W^+$ in the first order in weak
interactions. The corresponding width is well known 
\be {\Gamma_t\over m_t}={\alpha_w \over 16} ({m_t^2 \over m_W^2})(1+2 {m_W^2 \over m_t^2}) (1-{m_W^2 \over
  m_t^2})^2 \ee
  However, for both $t$ and $W$ bound in the bag it is by no means certain whether
 $m_t>m_W$ or not. As we had shown in \cite{topbags},
 for large enough bags the lowest top levels are in fact
$lower$ than the lowest $W$ ones.

When the corresponding   top energy levels
 are still above the W states,
 such weak reaction is  possible. Crude estimate
of the weak decay width can be obtained if in this formula
we substitute masses by energies, writing (for $E_t>E_W$) the width
as
\be {\Gamma_t\over E_t} \sim 10^{-3} (1-E_W/E_t)^2 \ee
In practice, this leads to a lifetime of the order of $\tau_{weak}\sim m/\Gamma\sim 10^3$. 

When the  top quarks are heavier than $Ws$ in the bag 
 decays  $t\rightarrow  Wb$  become impossible. The
weak decays would the proceed into the 3 body, say $t\rightarrow b l\nu_l$ 
like in the usual beta decays.
This produce an extra power of $\alpha_w$ and much stronger dependence
on the energy released  because of 3 body phase space. This width, crudely estimated as 
\be {\Gamma_t\over E_t} \sim \alpha_w^2 ({\Delta E_t\over E_t})^5  \ee
where by $\Delta E_t $ we mean the change in the total energy
when one top disappears, is way too small and exceed all the scales in the problem (except the time of cosmological expansion).

In summary, $m\tau_t\sim 200$ is the lifetime due to to strong top annihilation, while their weak decays  take no less than $mt\sim 10^3-10^4$ time, or even much more if the bags are large enough. This should be
compared to $\tau_W m$ of about 100 for the $W$'s, which would clearly be shorter and thus limit the lifetime of the bags.
The lifetime $\tau m$ of about
20 for the  bosonic spots in simulations is due to thermal evaporation: presumably a deeper binding and decrease of internal temperature  due to tops will improve it.
We conclude with a crude estimate of the bag lifetime as $\tau m\sim 100$.

\section{The sphaleron transitions in the top-stabilized bags}
\label{sec_sphal_topbags}
\subsection{The  exponential effects and the lifetime}
As argued before, tops would be an important ingredient in the metastable bags with near-zero Higgs VEVs. Now we turn to the next question:
how much the {\em sphaleron rate} can be affected by their presence, as compared to purely bosonic ones in
the simulations \cite{GarciaBellido:2003wd}? 

Since in this case there are more particles in the bags, with the tops added to gauge bosons, one may naively think that
the bag energy would grow. But using a simple bag model we have already pointed out that this would not be the case:
the mechanical stability require total pressure to be zero, which related the internal pressure to the bag constant $B$. 
No matter which light degrees of freedom are inside, the bag energy would scale as $BV$.
The Boltzmann factors for sphalerons however would still depend on  the number of degrees of freedom $N_{DOF}$ 
in the bag because the temperature depends on it. 
This again follows from the mechanical stability condition
 \be N_{DOF} T_{in}^4 \sim B  
 \label{eqn_stab}\ee
 Thus adding tops to the bag simply reduces the internal temperature. 
Returning to
  the exponential approximation for the sphaleron rate (\ref {eqn_cos_rate}), we now put into it the optimal radius
  (\ref{eqn_optimal_size}) and also rescale the temperature as in  (\ref{eqn_stab}).
   One finds that the exponent of the sphaleron rate depends on the number of degrees of freedom as

 $\sim   (N_{DOF})^{0.25}$
and  changes by just few percents only.
 We thus conclude that 
 the cancellations due to mechanical stability condition  (\ref{eqn_stab}) makes the effect of tops on the exponent small,   within the level
of uncertainties of the model used.

 Now we turn to the discussion of the preexponent. Obviously it depends nontrivially on the presence of the tops in the bag, as fermions
 are related to the probability of topological fluctuations by the anomaly: we will turn to this effect in the next section. 
 If the 
 $T^4$ in the rate comes from some thermal average,  one may think that it will come multiplied by (at least) one power of $DOF$ factor.
If so,  again the increase of the $DOF$ in the bag due to tops would be partially canceled out by a reduced internal temperature, since the product is the bag pressure $B$.
The preexponent simply has to be recalculated, for COS sphalerons in the bag environment (in which many fields such as gluons are absent because of
different cosmological scenario).

We have argued above that tops have longer lifetime in the bag, which 
may increase the overall bag lifetime. At the other hand, bag stability relies on $W,Z$ presence and thus it cannot exist
longer than the $W$ lifetime, $\tau_W m\sim 200$. Thus the potential enhancement
of the hot spot lifetime cannot be larger than factor 10.  


\subsection{Recycling the top quarks}

  The well known Adler-Bell-Jackiw anomaly  require
 that  a change in  gauge field 
topology  by $\Delta Q\pm 1$ must be accompanied by a corresponding
change in baryon and lepton numbers, B and L. More specifically, 
such topologically nontrivial fluctuation
 can thus be viewed as a ``t'Hooft operator" 
with 12 fermionic legs. Particular
fermions depend on orientation of the gauge fields in the electroweak SU(2): since we are interested in
utilization of top quarks, we will assume  it to be ``up". In such case the 
produced set contains $t_rt_bt_g c_rc_bc_g  u_ru_bu_g \tau\mu,e$ , where $r,b,g$ are quark colors.
We will refer below to this process as the  $0\rightarrow 12$ reaction.
Of course, in matter with a nonzero fermion
 density  many more reactions
of the type $n\rightarrow (12-n)$  are allowed, with $n$ (anti)fermions captured from
the initial state.

Although in this work we try to follow thermal language  (local kinetic equilibrium) as much
as possible, let still discuss what should be done
in a purely dynamical out-of-equilibrium setting.
In principle, one should proceed  quantum mechanically,
starting and ending 
with certain number of ingoing and outgoing quark and gauge quanta in certain in and out states (e.g. fixed momenta)
projecting those into  the fermionic and gauge field configurations of the semiclassical theory
of the sphaleron process. The sphaleron
itself is just one  -- saddle -- point at $t=0$ 
on the path, which start from convergent waves and end with divergent ones.

 The whole 
classical solution describing the expansion stage at  $t>0$ has been worked out for COS sphaleron explosion,
and for the ``compression stage'' at $t<0$ one can use the same solution with a time reversed. 
  At very early time or very late times
 $t\rightarrow \pm\infty$  the classical field become weak and
describe convergent/divergent spherical waves, which are nothing but
certain number of colliding gauge bosons. 
 Fermions of the  theory should also be treated accordingly.
  Large  semiclassical parameter -- sphaleron energy over temperature -- parametrically leads to the
 assumption that total bosonic energy is much larger than
that of the fermions, so one usually ignores backreaction and
consider Dirac eqn for fermions in a given gauge background. For KM sphaleron and effective $T$ we discuss,
this parameter would be $\sim 70$, which is indeed large compared to 12 fermions. However in the case
of COS sphaleron we are going to use the number is about $\sim 10$,  comparable to the number of fermions produced.
It implies that backreaction from fermions to bosons is very important.

To our knowledge, the only
(analytic) solution to 
Dirac eqn  of the ``expansion stage'' 
was obtained in \cite{Shuryak:2002qz}, it describes motion
 from the sphaleron zero mode at $t=0$ and
all the way to large $t\rightarrow+\infty$, the outgoing physical 
state of fermions, with momentum distribution
(\ref{eqn_q_spectrum}). A new element we are adding now
is  that its
time-reflection  can
 also describe the compression stage,
from free fermions captured by a  convergent
 spherical wave of gauge field at  $t\rightarrow-\infty$
 and ending at the sphaleron zero mode at $t=0$. 

\begin{figure}[t]
\centering
\includegraphics[height=4cm]{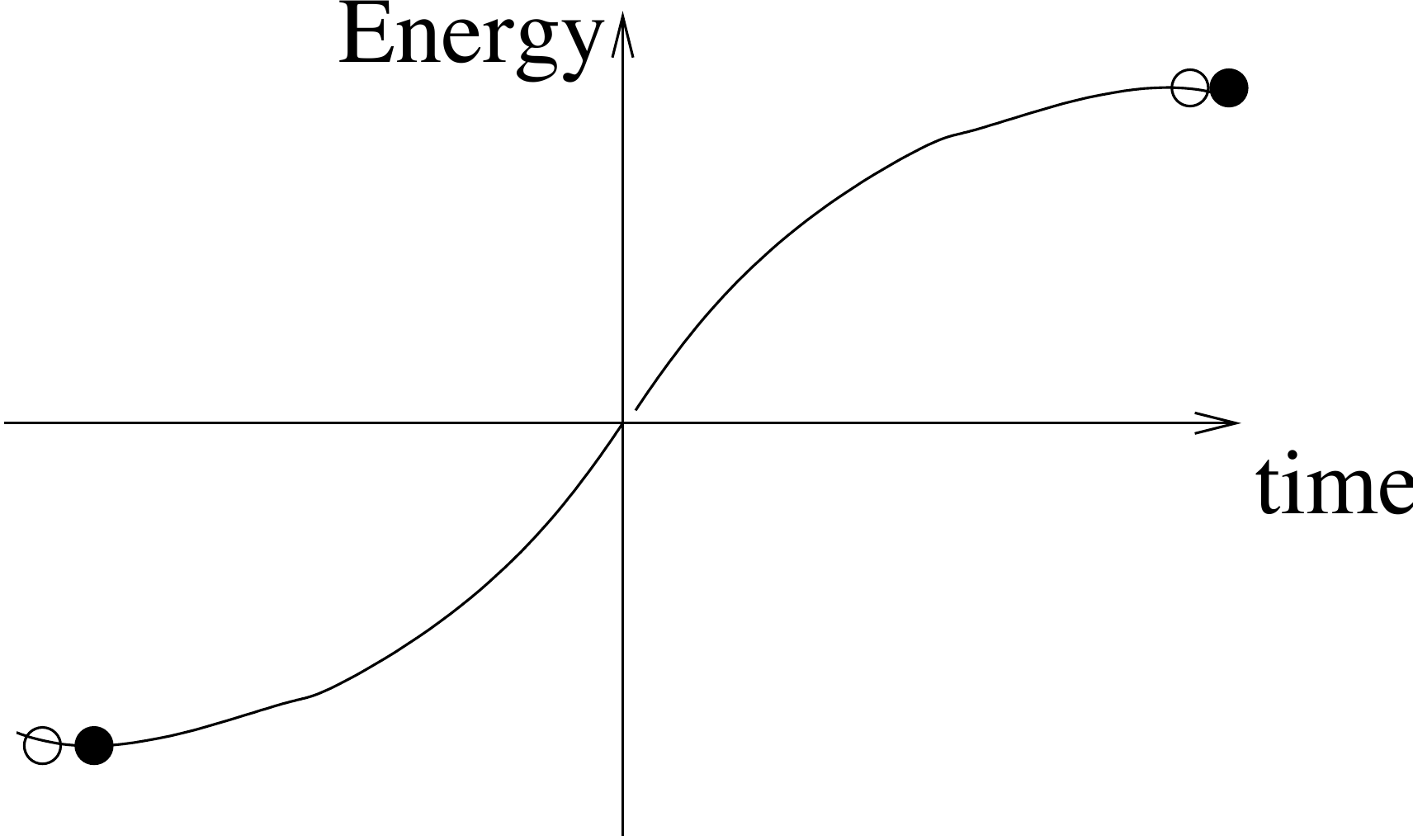}
\caption{Schematic picture of fermionic level motion as a function of
  time. Open and closed circles indicate unoccupied and occupied
states.  }
 \label{fig_levelmotion}
\end{figure}

 How the
corresponding fermionic levels move is schematically
 shown in
Fig.\ref{fig_levelmotion}. In
the Dirac sea notations, the level moves from some
negative initial  energy at early time to positive one at late time.
For COS sphalerons  the mean energy is $E_q=-3/\rho$ -- to  $E_q=+3/\rho$ By symmetry, the level is
crossing zero  at $t=0$: the corresponding fermionic
zero energy state is known as the sphaleron zero mode.
 If the level was occupied
(as most of levels below
the surface of the Dirac sea are) it remains occupied and one gets
new fermions produced. If there was antiquark (an unoccupied
state or a hole, indicated
by the open circle) it will remain unoccupied, thus having
no meaning after the level moves to positive energy, where
most levels are unoccupied.  
So, antiquarks -- holes at negative energy sea -- 
  are actually  $decelerated$ by the
radial electric field
 to zero energy  modes, their energy being transmitted to
the gauge field.

  This conclusion is very important: it implies that energy
of the initial fermions can be incorporated and used
in the sphaleron transition. If initially fermions were
in certain state 
(e.g.  certain level of the Higgs bag $| in >$ ),
the fermion part of the sphaleron amplitude would include
the amplitude of their capture into  
 a ``sphaleron doorway" state $|door\,way>$  by a convergent gauge
 field wave. As we will discuss in detail below,
we will find that the optimum way
to generate sphaleron transition is to use 3 initial top quarks,
considering the $3\rightarrow 9$ fermion process instead of 
the original  $0\rightarrow 12$ one.
In order to satisfy Pauli
principle and fit into the same sphaleron zero mode, colors of the 3
quarks of each flavor should all be different.

Let us show by simple estimates that under conditions we are discussing the  $0\rightarrow 12$  fermion production process
is actually impossible. 
  The final kinetic energy for fermion produced by COS sphaleron is
$E_q= 3/\rho$ which is $\approx 200\, GeV$ for the $m\rho=3.9$
example displayed as typical in Fig.\ref{fig_profile}.
  Multiplied by
  12 fermionic species produced it would require $2.4 \, TeV$ of
  energy which apparently
 exceeds the total available energy
of the gauge fields in the topological fluctuations observed numerically.

  (Note finally, that inclusion of the CP-violating processes -- such as induced by CKM matrix -- may lead to a small difference
between recycling  of the top and the antitop quark amplitudes. This asymmetry would then be
a mechanism for the baryogenesis.)

 The $3\rightarrow 9$ fermion process saves a lot of energy,  
as in it  the initial top quark energy can be completely
 transferred 
from the ``sphaleron doorway state'' to the gauge field
during the compression stage. In the example we discuss,
with $m\rho=3.9$, this mean  energy of 3 top quarks is 
\be \Delta E=3*3/\rho\approx 600\, GeV\ee
An estimate with exponential accuracy  we find its
enhancement by the factor of about \be F_{recycling}\sim  \exp(\Delta E/T)\sim 20 \ee
using the hot-spot temperature $T_{in}$.

Note that the same amount of energy
 should also $not$ be produced in the final state:
 thus in terms of energy balance this amount is doubled, saving $\sim 1200 \, GeV$
 and making the transition possible. 
  Yet this gain does not increase the rate any further,
 as the system  falls downhill classically the probability of everything to happen being just 1,
 due to unitarity.

Can one continue to use the same mechanism for more initial-state
fermions, for example considering the capture
of 3 tops plus $n$ other quarks, gaining another factor $\sim e^{n\Delta E/T}$? In particular, should one focus on $6\rightarrow 6$ transition, which is ``energy neutral''?
The answer to this question is $negative$:
for quarks other than $t$ there is no reason to get trapped inside
the no-Higgs spots, as Higgs effect on them (the mass) is too small 
compared to even low bulk $T_{out}$ and they occupy
the bulk of the volume. Roughly  their  density
is thus smaller than that of top quarks by the factor
$(T_{out}/T_{in})^3$, which is few percents.  The cube of this factor obviously
cannot compensate the gain from the exponent.

The preexponent for the
 $3\rightarrow 9$ fermion process should include 
projection of  the initial states of fermions (say bound states of 
 $\bar t\bar
t\bar t $) in a  bag onto the appropriate
``fermionic doorway state'' of the sphaleron process.
As it was already mentioned, 
 those have been found 
  in
\cite{Shuryak:2002qz} for COS sphalerons: they have simple exponential form. Lacking bound state
wave functions in momentum representation, we can use the convolution with thermal state

\be  P_{\bar t}={1\over2 n_t(T_{in})}\int_0^\infty dp {(2p\rho)^2 exp({-2p\rho}) \over exp[(p-\mu_t)/T_{in}]+1} \ee
Factor 2 in denominator
appears is because we only integrate over momenta directed inward,
and $n_t(T_{in})$ is the thermal quark density.
Note that  $m\rho \sim 4$ in numerical model we discuss, so the corresponding effective temperature of the doorway state
is $1/(2\rho)\sim m/8\sim 30 GeV$ is softer than the actual internal $T_{in}\sim 200\, GeV$. This mismatch would somehow reduce
the preexponent. On the other hand, one can further optimize the rate, selecting optimal (higher than average)
momenta of the top quarks, to affect even the exponent.

\section{CP violation} 
Finally, let us discuss the last crucial component of the baryogenesis, the CP violation. 
As it is widely known, all so far observed CP violation in $K,B$ meson decays can be explained inside  the SM, in which it comes from the phase of the CKM matrix.   

(The only exception to this is a very recent result from D0 collaboration \cite{D0_CP} on same-sign dilepton asymmetries 
which are about -0.01,   a 3.2 standard deviation  from much smaller SM prediction   $(-2.3  \times 10^{-4}$. Its origin is believed to be the $B_s \leftrightarrow \bar{B_s}$
oscillations: in a couple years we will know if it withstand further data samples from D0,CDF and LHCb. ) 

 Returning to SM and its CKM matrix, it would be desirable to establish a specific process and prediction, which will follow from it for the cosmological scenario 
 under consideration.  Since in this scenario 
 two (of the three)
Sakharov's conditions -- the deviation from equilibrium and the baryon number violation -- are in a sense
maximized,  so the required magnitude of the CP violation in this scenario need to explain the observed asymmetry is minimized, to something
of the order of $10^{-7}$ or so.  Below we discuss whether such level of CP violation can actually be reached inside the SM.
   
 \subsection{Optimizing the one-loop CP-odd effective action }
  
The magnitude of the CP violation can be derived by standard  one-loop effective action which is $log det D$, with the Dirac operator $D$
including Higgs and gauge fields. The well-known Jarlskog argument corresponds to the situation in  which the field mass scale \be F^2\sim W_{\mu\nu} \gg m_q^2\ee 
is larger than that of any quarks. In this case the coefficient of the effective operators must be proportional not only to the so called Jarlskog
invariant $J$ but also to extremely small Jarlskog determinant
\ba  \delta L_{CP}  \sim {J\over F^{8}} (m_t^2-m_c^2)(m_t^2-m_u^2)(m_c^2-m_u^2)  \nonumber  \\ (m_b^2-m_d^2)(m_b^2-m_s^2) (m_s^2-m_d^2)
\sim J m_t^4 m_b^4 m_c^2 m_s^2 /F^{12}\ea
because when each two same-charge quarks are degenerate the CP odd phase can be rotated away to zero, nullifying the Cp-odd effect.
Since the field inside the bags we discuss is of the scale $F\sim 100 GeV*\sqrt{N}$, this is indeed the case and thus we have to conclude
that CP odd effect $inside$ the  bag is too small.

Let us now try to optimize the effect, by looking at all scales $F=m_t,m_b,m_c,m_s$, subsequently. The corresponding expressions and numerical values 
(for $J=10^{-5}$ and actual quark masses) are
\ba  \delta L_{CP}(m_t) \sim {J\over m_t^{4}} (m_c^2-m_u^2)  (m_b^2-m_d^2)(m_b^2-m_s^2) (m_s^2-m_d^2) \nonumber \\
\approx 3 *10^{-13} \, GeV^4 \,\,\,\,
\ea

\ba  \delta L_{CP} (m_b) \sim J (m_c^2-m_u^2)  (m_s^2-m_d^2)\approx 4 *10^{-7} GeV^4
\ea

\ba  \delta L_{CP} (m_c) \sim J m_c^{2}  (m_s^2-m_d^2)\approx 4 *10^{-7} GeV^4
\ea

\ba  \delta L_{CP}(m_s) \sim m_s^4 J \approx 4 *10^{-9} GeV^4 \ea

Thus, at this level of accuracy, the scales of $b$ and $c$ give comparable effects. 
There are however two more consideration which favor the $b$ quark scale
as the optimal one. 

One is related with ``thermal masses" in the medium
 of the quark $outside$ bags. Weak thermal (Klimov-Weldon)  quark mass for left handed
fermions due to scattering on thermal W,Z  is (see e.g. \cite{FarrarShaposhnikov}) 
\be \delta M^2= {\pi \alpha_w\over 8} T^2 \ee 
  which for the outside temperature $T\sim 50\, GeV$ is about
$(5 \, GeV)^2$.
 This independently points to  the possibility
to go down to the  scale  of  the b quark mass in all fermionic propagators.  
  
  The second consideration is specific to the hybrid scenario we discuss.  In this case the light quarks are produced not from Higgs but from weak $W,Z,t$ decays,
  and gluons from quark annihilation: and at the initial period of the electroweak phase under consideration both are not happened yet. 
  For this reason we have not included gluonic mass in the previous paragraph. For the same reason we expect the mechanism due to a
  difference in $s$ and $d$ quark scattering  on the ``bags"    be  subleading to $b$ scattering or $b-t$ transitions.
  
  In order to do the actual calculation one needs to calculate explicit CP violating Lagrangian and integrate it over the space-time volume. 
 As a crude estimate, let us add the minimal power of the coupling $(4\pi\alpha_w)^2\sim 0.2$ needed for 4 CKM matrices,
  and ``coherent space-time volume"  $V_4$ outside the spot in which the gauge field is  at the scale of $m_b$, times the estimated  $\delta L_{CP} (m_b)$.
 One can then conclude that for $V_4\sim 1\, GeV^{-4}$ the total effect may reach the needed $10^{-7}$. 
 Although this volume is comparable to space between the ``spots" times the lifetime, 
 it is highly nontrivial that the ``coherent space-time volume"  (in which the effect maintains the same sign)
  can be that large, as the structure of the Lagrangian 
 is complicated and must include the 4-dim epsilon symbol. Needless to say, the corresponding Lagrangian at the $m_b$ scale is not yet calculated (to our knowledge).

 The closest to it is the calculation in Ref.\cite{Hernandez:2008db}  for the scale below $m_c$, which produced the following
 operator of
dimension-6 
 \ba \delta L_{CP}(<m_c)
  \sim J m_c^{-2} \epsilon^{\mu\nu\lambda\sigma} \\ \nonumber [Z_\mu W^+_{\nu\lambda} W^-_\alpha (W^+_\sigma W^-_\alpha  + W^-_\sigma W^+_\alpha)+cc ]\ea
 Some elements of its general structure, with 4-dim epsilon (providing negative P parity) and
four $W$ vertices, providing the Jarlskog invariant $J$ from four CKM matrices, are  generic. This structure alone implies quite
complicated product of the fields, with a necessary nonzero electric field. 

Averaging this operator over fields found in numerical bosonic simulations has been performed in ref.\cite{Tranberg:2009de},
with the result  four orders of magnitude $larger$ than needed. It is  however clear  that  one actually cannot apply this Lagrangian,
 since this operator was derived for a completely different field scale $F\ll m_c$, which is not the case in 
the numerical simulations in question. Not surprisingly, the obtained result is 
larger than the observed one: the dimension 6 of the field would produce very large contribution from the ``hot spots" with very strong gauge fields.
However, as follows from Jarlskog argument (repeated above)   at larger field there is additional strong suppression due to mass differences, which will
kill it as detailed above.

\begin{figure}[h]
 \includegraphics[width=8cm]{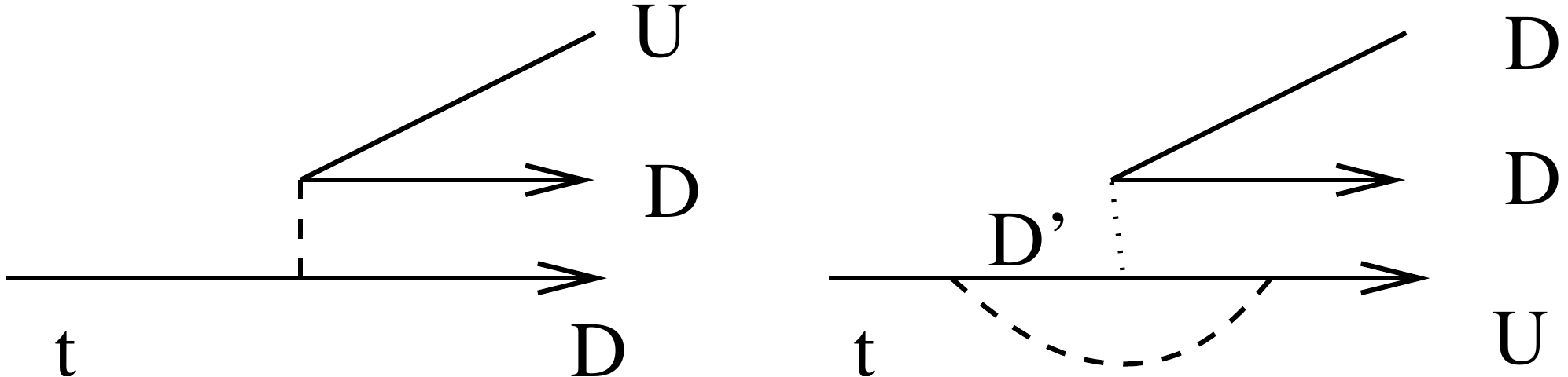} 
\caption{Two interfering diagrams in top quark decays: the dashed line is the $W$ and the vertical dotted line in (b) is the gluon. $D=b,s,d,U=c,s$ are generic
names for decay products, $D'$ is the intermediate quark in the loop which is summed over down flavors.}
\label{fig_tdecays}
\end{figure}

\subsection{CP violation in top decays:  the ``penguins" and the final state interaction}

It is well known that CP violating effects in particular meson decays are much larger than what one would get
from the one-loop gradient expansion effective Lagrangians like the one discussed above.  Instead of high order in weak interaction, 
 one may find the needed four CKM matrices  in the interference terms between lowest order weak interaction and  strong radiative corrections, 
 the so called ``penguin" diagrams.
Fig.\ref{fig_tdecays} shows two generic diagrams, which can interfere and produced the CP-violating asymmetries from the famous complex phase in the
CKM matrix. As seen from the figure, there are six generic decays for all possible values of the final quarks:
\be t\rightarrow \bar{d}du,  t\rightarrow \bar{s}sc,  t\rightarrow \bar{d}dc,    t\rightarrow  \bar{b}bc ;  t\rightarrow  \bar{s}su;  t\rightarrow  \bar{b}bu \ee
The amplitude of the first one can be written as
\be A( t\rightarrow \bar{d}du) =V_{td} V_{ud}^*+ V_{tb} V_{ub}^*F_{penguin} \ee
in which we selected $D'=b$ and introduce the so called penguin factor, including in particular
the propagator of an extra gluon and the strong coupling constants.
 The interference term between these two terms would provide the needed four CKM matrices,
while the ``time arrow" can be provided by a specific direction of the flow, e.g. leakage of bound tops into the unbound $lighter$ quarks leaving the bag. 
(Note again, that in equilibrium there would be inverse processes which will cancel the asymmetry, but those are absent
in the scenario we discuss: the light quarks are very much out of equilibrium and their density is negligible at the time we consider.)

The ratio of the two terms generates CP asymmetry, which for this decay is as large as 
 \be CP\sim {V_{tb} V_{ub}^*F_{penguin} \over V_{td} V_{ud}^*} \sim  \eta F_{penguin}  \ee     
Other modes have smaller asymmetry by powers of $\lambda$, but their decay rates are larger.

To get the average effect one has to evaluate the effective CP-odd Lagrangian or
 the imaginary part of the product of the two terms, their interference.  For all 6 decays mentioned above
\ba {\delta L_{CP} \over F^4} \sim \alpha_w^2 Im[V_{td} V_{ud}^* V_{tb}^* V_{ub}] F_{penguin} \nonumber \\
\sim \alpha_w^2 \eta \lambda^6 F_{penguin}  \approx 10^{-8}   \ea
where we have used the Wolfenstein parameterization (to $\lambda^4$ accuracy) of the CKM matrix, and include ``penguin suppression"  factor $\sim 1/3$.
The net effect of this Lagrangian would be a
difference between the number of bound tops and antitops in the bag, and with ``top recycling" leading to asymmetric sphaleron transitions and
baryon asymmetry. 

(Note that the quark mass differences  entering the Jarlskog determinant argument of the previous subsection.
 It resigns in the summation over the $D'$ quarks. If all three Penguin diagrams would be exactly the same, one would get the following combination
 \be  V_{tb} V_{ub}^*+ V_{ts} V_{us}^* + V_{td} V_{ud}^* \ee
which is zero because of the orthogonality condition of the two rows. Here however where different quark  masses and  ``external conditions" (in form of the
external thermal mass in the propagators of the $D'$ quarks) should help to produce different coefficients.
 Since the scale of such external effects are of the order of $F\sim$ few GeV, they clearly separating the $b$ penguin term from 
two others. As in the previous section, the remaining suppression of the order of $(m_s-m_d)/F$ may yet appear. )
 
 Most CP in meson decays are due to mixing of the neutral meson-antimeson states: but this is not needed. Top/antitop decays are like
 those of the charged meson, which will have a CP effect due to
the final state interaction, or ``strong phases" as they are usually called. The first and second diagrams should be ``dressed" by the gluon
 exchanges between the quarks (not shown in Fig.), it will result in some phases $\delta_{direct},\delta_{penguin}\sim O(\alpha_s)$, which are the same
 for quarks and antiquarks. They are not the same for different final states, for example if the quark pair produced from a gluon in a penguin has small
 invariant mass, it corresponds to a pair produced at relatively large distances from the original quark, while the ``direct" decay diagram is near-pointlike because
 of large $W$ mass. Standard arguments show that there is a difference in the decay rates of particle and antiparticle
 \be |A(t\rightarrow f)|^2-  |A(\bar t\rightarrow \bar f) |^2 \sim sin(\delta_{direct}-\delta_{penguin})sin(\delta_{CP} )\ee   
 Only recently the first such asymmetry has been observed in a difference between the width
  $\Gamma(B^+     \rightarrow      \rho_0 K^+)$ decay and that of its CP conjugate.
 So, there is no doubt that $partial$ widths of the $t$ and $\bar t$ decays can differ. 
 While the CPT theorem forbids the difference in their total width, for  {\em time dependent} bags well out of equilibrium 
one may still hope that  it would be possible to go around strict CPT it and generate a difference in $t$ and $\bar t$ population,
perhaps even  of the magnitude comparable to the observed asymmetries in its decay modes.

\subsection{Baryon charge separation during the bags formation}
Another possible manifestation of the CP-odd phenomena is possible separation of the quarks vs antiquarks already during the process of bag formation,
in the initial very off-equilibrium stage. If it happens, then the amount of tops and antitops in the bag would already be different, leading to an asymmetry
in sphaleron transitions which ``recycle" those tops, as detailed above.

Farrar and Shaposhnikov   \cite{FarrarShaposhnikov}  were the first who discussed it in detail, in a specific case of strange quarks and near-equilibrium 
bubbles at $T=Tc$, assumed at the time to be the first order transition.  Although we would use this phenomenon in a quite different setting, we would still call
it ``the Farrar-Shaposhnikov (FS) mechanism". Its essence is the interference between the (CP-even) scattering/reflection on the bag wall
and the (CP-odd) transition via the CKM matrix into a different flavor, reflection, and return to the original flavor. Let us remind the reader how it works.
Let us start with flavor $f_1$: if it is reflected from the bag wall it  gets the phase $\delta_1$. Alternatively it can turn into the flavor 2 and scatter with the phase   $\delta_2$,
picking on the way CP-odd phase from complex CKM matrix elements. The total baryonic current is a difference between those for quarks and antiquarks
\ba \left| e^{i\delta_1 } + C e^{i\delta_2 +i\delta_{CP}} \right|^2-\left| e^{i\delta_1 } + C e^{i\delta_2 -i\delta_{CP}} \right|^2 \nonumber \\
\sim sin(\delta_1 -\delta_2) sin(\delta_{CP})
\ea
The essential part of the argument is that two flavors under considerations should have different scattering on the bag, as the effect disappear at $\delta_1=\delta_2$.
This condition is especially difficult for light quarks, as they only weakly interact with the Higgs and thus the bag. FS has argued that the effect should be nonzero for
$s$ relative to $d$ quarks, but only in a narrow strip of energies near the level crossings, where $s$ quarks are all reflected (large $\delta_1$) while $d$ are
transmitted (negligible $\delta_2$).  As a result,  their estimate for the FS effect is proportional to a small parameter corresponding to the relative width of this region, $(m_s-m_d)/T\sim 10^{-3}$. 

Another important part of their paper (which we would not remind here) is discussion of the unitarity and CPT constraints on possible effects, 
leading to a standard conclusion that in equilibrium (e.g. for a bubble wall at rest with the medium) the effect  vanishes, as it indeed should from general considerations
\cite{Sakharov}.

Let us now point out large differences which exist between the matter properties discussed by FS and that in the hybrid scenario under consideration. First of all,
there are drastic differences in its chemical composition. Equilibrium matter at electroweak scale has about hundred effective degrees of freedom, including
all quarks and leptons, as well as gluons. Therefore effective fermionic thermal masses are dominated by the gluonic part, with a scale $\sim T/2\sim 50\, GeV$.
In the hybrid scenario there was not yet enough time to generate light quarks, leptons and gluons: the only quanta produced are weak gauge bosons $W,Z$, Higgs  and
  the top quarks, all due to their large coupling to (strongly fluctuating) Higgs field. The spatial distribution of tops is very space and time-dependent, as
  they flee the regions with large Higgs VEV and are collected into the bags we discuss. While doing so, they can also migrate, via the CKM matrix, into all other flavors,
  driving the FS mechanism and producing different top and antitop population in the bags. 
   
   For simplicity, let us imagine the bag to be very large and its boundary flat, separating the space into two halfs, with symmetric ($v\approx 0$) and broken ($v\approx v_0$)
   phases. The flow, into symmetric phase, is driven by the difference in the chemical potential. As we have discussed throughout this paper, the temperature is also 
   not constant over the divide, with $T\sim 50 \, GeV$ in the broken phase outside the bags. Tops in the bags (symmetric half) with energy less than the top mass outside  $E<M_t(T_{out})$  cannot leave the bag and are reflected back if collide with the boundary: thus large $\delta_t$. If they however migrate into the $c,u$ quarks, those
   hardly see a bag at all, thus $\delta_c,\delta_u\approx 0$. Additional small  parameter appearing in FS estimate is the part of the energies
   in which $c,u$ contributions do not cancel is $(m_c-m_u)/T \sim 1/50 $, not so small.  Together with Jarlskog $J$ from CKM, it will generate asymmetries of the order
   of $10^{-6}$ or so, which is right in the needed ballpark.   
   
(A word of warning:  The original FS estimate of the effect has been criticized in refs \cite{FS_critics}. One of the points is that additional suppression appears from the widths of quarks coming from in media scattering, and we are not sure what is the final word in the polemics about it. Anyway, this criticism is much less relevant to the  scenario discussed, as larger scale $m_c$
is substituting $m_s$ in FS, which is rather close to -- much reduced relative to equilibrium matter -- rescattering widths.) 

\section{Summary and Discussion}
In summary, we have considered the role of the $W-Z$-top bags in baryogenesis. We
have used,  as a specific example,  numerical simulations
for the hybrid preheating scenario from Ref.\cite{GarciaBellido:2003wd}.  

(i) We first reproduced the main results of  those simulation using the $W-Z$ bags and COS sphaleron, including qualitatively the sphaleron 
size and rate found numerically. 

(ii) We then discussed the role of the top quarks and  found them to be quite important. They are produced  via
$HH\rightarrow \bar{t}t$ more effectively than W. They are collected into the bags together with the gauge quanta, presumably 
producing bags of  larger size, with improved  lifetime. We 
however found that the mechanical stabilization condition
(which  makes bags with tops cooler than pure W-bags)
basically makes the sphaleron exponent constant, independently of how many top quarks are in the rate.

 (iii)
In addition, there is an interesting effect of ``top quark recycling".
This effect helps 
the transition, providing significant ($\sim 1/3$) fraction
of the energy needed to reach the sphaleron mass. The recycling of 3 tops increases  the rates by  another factor 20.

(iv) We then provide arguments that optimal CP violation should happen either due to high order weak interaction
in the area 
 between the bags, at a scale of $m_b\sim 5\, GeV$,
or in top decays, or in top selection during bag formation due to Farrar-Shaposhnikov effect. Although we are not able to provide specific calculation of any of the
three possibilities at this time,  crude estimates  show that any one of them can potentially provide a magnitude of $10^{-7}$ which, 
  together with the estimated sphaleron rate,
   would generate the baryon asymmetry
of the observed magnitude from known CKM-base mechanism alone. (If the D0 result on much larger CP violation in semileptonic $B_s$ decays
would stand in time, one should of course also look at its origins as well.)

 Let us finish this paper with a historical analogy. We do see stars on the sky which shine: yet basic nuclear physics  was not enough
 to explain why they can do so, falling short by many orders of magnitude. It took decades of patient and specific work to uncovered the nontrivial ways
in which the star cycles work. Those crucially depend on some catalysts, whose presence is very small but crucial.
 Perhaps the $W-Z$-top  bags are also a kind of a catalyst, or a doorway state, helping to increase the probability
 of the baryon number violation  and of the CP violation.

    \section*{Acknowledgments}
      
      The work of VF  was supported by the Australian Research Council and NZ 
Marsden fund. The work of ES is supported by the US DOE grant DE-FG-88ER40388. Both of us are thankful to the
New Zealand Institute for Advanced Studies for kind invitation, during which some of this work has been done.
We would also thank our collaborators Marcos Crichigno and Michail Kuchiev on other parts of the project, for helpful discussions.

\appendix
\section{The coupling constants and parameters}
Let us remind the reader
the hierarchy  of the coupling constants, as it reflects the order in which 
various processes should be considered.

 The largest coupling
constant is still that of strong interaction, \be {g_s^2\over 4\pi}= \alpha_s(100\, GeV)\approx 0.11  \ee
but all the original bosonic fields of the electroweak theory $W,Z,H$ can only interact with
strong interaction via quarks, so it has to play so to say the secondary role.

Quite  close to it  is the Yukawa coupling of the top quark to Higgs, if written in the same form
it is
\be  {g_{ttH}^2 \over 4\pi}=\alpha_t\approx 0.08 \ee
and this would be responsible for most of the effects we will discuss below.
Other Yukawa couplings are much smaller and can be neglected: for example for the next b quark
such combination is smaller by the factor $(m_b/m_t)^2\approx  7.4*10^{-4}$.

Weak interaction coupling is naturally smaller
\be \alpha_w\approx .03 \ee
in the standard model: note however that in numerical simulation we discuss a much smaller
value has been used, presumably for some methodical reasons. We will not
include weak interactions except in top quark decays. 
Like in many other works, we will ignore electromagnetism (the U(1) gauge part),
putting the Weinberg angle to zero: thus $m_W=m_Z$.

 It is well known, the Higgs mass (or self-coupling $\lambda$)
 remains the last unknown parameter of the SM. 
 Our previous studies of the top-balls put the Higgs mass to a round number
 $M_H=100 \, GeV$.
 
 However, since we try to explain numerical bosonic simulations \cite{GarciaBellido:2003wd}, let us mention
that their choices for the  Higgs mass and VEV are different. 
Details can be seen in their original work, for  
 example  the particular set of parameters, called A1 in \cite{GarciaBellido:2003wd} for which the
 pictures we used are based have rather heavy Higgs mass
 \be \lambda=g^2/2=0.00675, \, \, m_H/m_W=4.65\ee
and also a different the value of the electroweak coupling is $g_w =1/20$.
The Higgs VEV is still $v=246\, GeV$ as in real world.  
The mass parameter of the Lagrangian
related to physical Higgs mass by $m=m_H/\sqrt{2}$ is  chosen 
as a basic mass/energy/length/time unit,
which we will also adopt in what follows. With realistic $M_W$ it makes $m=264\, GeV$, not
far from $v$.

\section{Strong annihilation}

The differential cross sections for these two 
QCD reactions are well known, here we present them ignoring
 quark masses\footnote{Those cannot be ignored entirely 
because this simplification leads to IR
some divergences to be encountered below: yet those are logarithmic
and be regulated by mass when needed.}
:
\be  {d\sigma_{\bar t t\rightarrow \bar q q} \over dt}= {4\pi \alpha_s^2\over s^2}[1+u^2/s^2]\ee
 \be  {d\sigma_{\bar t t\rightarrow gg} \over dt}= {4\pi \alpha_s^2\over s^2} 
 ({t^2 +u^2 \over ut})({8\over 3}- {6ut\over s^2}  ) \ee
Integrated it we get\footnote{The second reaction, unlike the first, has logarithmic
divergence which can be regulated by a quark mass: the result however depend weakly
on the regulator and we keep the same coefficients for simplicity.}
 \be \sigma_{\bar t t\rightarrow \bar q q} \approx  \sigma_{\bar t t\rightarrow gg} \approx   {0.022\over s} \ee
Thus the contribution of those two processes to the decay rate  of the top quark can thus be written as
  \be   \Gamma_{\bar{t}t}=2*0.022 {  \int {d^3p_1\over  (2\pi)^3} {d^3 p_2 \over (2\pi)^3}  {f(p_1)f(p_2)\over 2p_1 p_2 [1-cos(\theta_{12})] } 
  \over \int {d^3p_1\over  (2\pi)^3}  f(p_1)  }
  \ee
where $f(p)$ is the thermal distribution and $\theta_{12}$ is the angle between momenta of two
colliding tops. Assuming that logarithmic angular integral is regulated by $(1-cos(\theta_{min}))\approx 0.1$ we get the following simple answer for the strong annihilation time
\be  \tau_{\bar{t}t}=1/\Gamma_{\bar{t}t} \approx 200/T\ee


\end{document}